\documentclass[fleqn,usenatbib]{mnras}
\usepackage{float}
\usepackage{newtxtext,newtxmath}

\usepackage[T1]{fontenc}
\DeclareRobustCommand{\VAN}[3]{#2}
\let\VANthebibliography\thebibliography
\def\thebibliography{\DeclareRobustCommand{\VAN}[3]{##3}\VANthebibliography}

\usepackage{graphicx}	
\usepackage{amsmath}
\usepackage{multirow}
\usepackage{multicol}
\usepackage{xcolor}

\newcommand{\teff}{$T_\textrm{eff}$}

\newcommand{\cd}{d$^{-1}$}
\newcommand{\luminosity}{$\log ({L/L_{\odot}})$}

\newcommand{\kms}{km\,s$^{-1}$}
\newcommand{\vsini}{$\upsilon\sin i$}

\newcommand{\tess}{{\it TESS}}
\newcommand{\kepler}{{\it Kepler}}
\newcommand{\dSct}{$\delta$\,Sct}
\newcommand{\gDor}{$\gamma$\,Dor}

\title[Asteroseismology of HD\,118660]{Asteroseismology of the mild Am $\delta$ Sct star HD\,118660 : {\it TESS} photometry and modelling}

\author[M. Sarkar et. al.]{
{Mrinmoy Sarkar}$^{1,2}$\thanks{E-mail: mrinmoysarkar@aries.res.in}, 
{ Santosh Joshi} $^{1}$, {Marc-Antoine Dupret}$^{3}$, {Otto Trust}$^{4}$, {Peter De Cat}$^{5}$, \newauthor{Eugene Semenko}$^{6}$,{Patricia Lampens}$^{5}$, {Aruna Goswami}$^{7}$, {David Mkrtichian}$^{6}$, {Drisya Karinkuzhi}$^{8}$, \newauthor{Ilya Yakunin}$^{9}$, {Archana Gupta}$^{2}$
\\
$^{1}$Aryabhatta Research Institute of Observational Sciences, Manora Peak, Nainital- 263002, India\\
$^{2}$M.J.P Rohilkhand University, Bareilly, Uttar Pradesh-243006, India\\
$^{3}$Space Sciences, Technologies and Astrophysics Research (STAR) Institute, University of Li\`{e}ge, B-4000 Sart Tilman, Belgium \\
$^{4}$Department of Physics, Mbarara University of Science and Technology, P.O. Box 1410, Mbarara, Uganda\\
$^{5}$Royal Observatory of Belgium, Ringlaan 3, B-1180 Brussels, Belgium\\
$^{6}$National Astronomical Research Institute of Thailand, Mae Rim, Chiang Mai-50180, Thailand\\
$^{7}$Indian Institute of Astrophysics, Koramangala, Bangalore-560034, India
\\
$^{8}$Department of Physics, University of Calicut, Thenhipalam, Malappuram 673635, India\\
$^{9}$Special Astrophysical Observatory, Russian Academy of Sciences, Nizhnii Arkhyz-369167, Russia\\ 
}
\date{Accepted XXX. Received YYY; in original form ZZZ}

\pubyear{2024}

\begin{document}
\label{firstpage}
\pagerange{\pageref{firstpage}--\pageref{lastpage}}
\maketitle

\begin{abstract}

We present the results of an asteroseismic study of HD\,118660 (TIC\,171729860), being a chemically peculiar (mild Am) star exhibiting $\delta$ Scuti (\dSct) pulsations. It is based on the analysis of two sectors of time-series photometry from the space mission TESS and seismic modelling.
It yielded the detection of 15 and 16 frequencies for TESS sectors 23 and 50, respectively. 
The identified pulsation modes include four radial ($\ell=0$) and five dipolar ($\ell=1$) ones. 
The radial modes are overtones with order $n$ ranging from $3$ and $6$. Such high values of $n$ are theoretically not expected for stars with the effective temperature of HD\,118660 ($\rm T_{\rm eff}\approx 7550 \rm K$ ) located near the red edge of the \dSct\ instability strip. 
To estimate the asteroseismic parameters, we have generated a grid of stellar models assuming a solar metallicity ($Z=0.014$) and different values for the convective overshooting parameter ($0.1\leq \alpha_{\rm ov}\leq 0.3$). 
We conclude that the analysis of the radial modes is insufficient to constrain $\alpha_{\rm ov}$ and $Z$ for \dSct\ stars. 
The value for the equatorial velocity of HD\,118660 derived from the seismic radius and the rotational frequency is consistent with values found in the literature.

\end{abstract}



\section{Introduction}

Asteroseismology uses stellar pulsations, which appear as periodic variations of brightness, radial velocities, and/or line profiles, to probe the internal structure and evolution of stars. Pulsating stars are ubiquitous in the Hertzsprung-Russell (H-R) diagram \citep{2022ARA&A..60...31K}.
The $\delta$\,Scuti (\dSct) stars are of particular interest for asteroseismology for several reasons. The richness of their pulsation spectra allows for the probing of several different internal layers. Their brightness makes them accessible for high-precision observations using both photometry and spectroscopy. 

The masses of \dSct\ variables typically range from 1.5\,M$_{\odot}$ to 2.5\,M$_{\odot}$ \citep{Breger_1979,2005JApA...26..249G,2021FrASS...8...55G,2022ARA&A..60...31K}. They are located in the H-R diagram where the classical instability strip intersects with the main sequence \citep{2004A&A...414L..17D,2005A&A...435..927D}, confined to the range of effective temperatures (\teff) nearly between 7000 to 9000\,K \citep{2011A&A...534A.125U}. 
In this region, stellar interiors change from radiative cores with thick convective envelopes ($\rm M\lesssim$\,1\, M$_{\odot}$) to the convective cores with thin radiative envelopes ($\rm M\gtrsim$\,2\,M$_{\odot}$). Hence, the study of \dSct\ stars can give important insights into the details of this transition.

Thanks to recent space missions like \kepler\ \citep{2008CoAst.157..266C}, {\it K2} \citep{Howell2014PASP..126..398H}, and the Transiting Exoplanet Survey Satellite (\tess; \citealt{2015JATIS...1a4003R}), we have a plethora of high-precision photometric data.
Analysis of these light curves shows that the \dSct\ stars mostly pulsate in pressure ($p$) modes with observed oscillation frequencies ranging from $\sim$5 to 100\,\cd\ \citep{2019MNRAS.485.2380M, 2019FrASS...6...40G,2020Natur.581..147B}. At the same time, the space-based photometric observations have revealed that a significant fraction of \dSct\ stars simultaneously exhibit frequencies that are typical for $\gamma$\,Doradus (\gDor) stars, making them hybrids \citep[e.g.][]{2014MNRAS.437.1476B}. 
The multi-periodic pulsational nature of \dSct\ stars make them an ideal test bed for asteroseismology.

The $\delta$ Sct stars pulsate in low-order radial and non-radial overtones. The corresponding observed amplitudes can be as low as a few tens of $\upmu$-mag (low-amplitude $\delta$ Sct stars; \citealt{murphy2014}) or as high as a few tenths of a magnitude (high-amplitude \dSct\ stars; \citealt{2008PASJ...60..551L}).
The opacity mechanism operating in the He\,{\sc ii} ionization zone is believed to be the primary source for driving pulsations in \dSct\ stars, although other mechanisms are likely to contribute \citep{houdek2000,antocietal2014,2020MNRAS.498.4272M}. 
Stellar metallicity is one of the key drivers in the excitation of their pulsations \citep{2018pas8.conf...59G}.

The influence of anomalous chemical abundances on stellar pulsations is not yet fully understood.  
Pulsating A-type stars, including the \dSct\ variables, share the same region of the main sequence with objects whose key feature is an abnormal chemical composition. These so-called chemically peculiar (CP) stars differ from their chemically normal counterparts as their spectra exhibit strengthened or weakened lines of specific elements \citep{1974ARA&A..12..257P}.
The CP stars constitute about 10--15\% of the population of stars with the same spectral type \citep{2009A&A...498..961R}.

The diffusion of atoms in stellar atmospheres is considered to be responsible for the formation of peculiarities in CP stars \citep{1970ApJ...160..641M, 2015ads..book.....M}. 
Some elements, particularly He, N, and O, settle in the atmosphere into the layers below the surface while others, such as Mn, Sr, Y, and Zr, are levitated out to the surface. 
When various mixing effects like convection, rotation, turbulence, or mass loss, are absent or do not play an important role, atomic diffusion can develop into a rather stable configuration with a non-uniform horizontal or vertical distribution of atomic species. A stable magnetic field can also provide the required stability. 
As a result, from an observational point of view, such a CP star may appear to have an abnormal chemical composition and, hence, metallicity. In reality, its average composition may not substantially differ from a non-peculiar star of the same origin and age.

The CP stars can be divided into smaller subgroups based on their most prominent chemical anomalies, binarity status, and magnetic properties. 
In this region of the H-R diagram, a star is classified as a magnetic CP star if it possesses a strong, stable, and globally organized magnetic field with a fossil origin, simple structure and surface strength of at least $\approx$300\,G.
Amongst the magnetic CP stars, a small fraction pulsates. They are known as rapidly oscillating Ap (roAp) stars because of their short pulsation periods ranging between approximately 5 and 30 min \citep{1982MNRAS.200..807K}.
Only about 100 of them are currently known \citep[and references therein]{Holdsworth2021MNRAS.506.1073H, Holdsworth2024MNRAS.527.9548H}.
The metallic-line (Am) peculiar stars are more numerous and are characterized by an underabundance of light elements (He, Ca, and/or Sc) in combination with an overabundance of iron-group metals.
Like \dSct\ stars, Am stars are generally considered to be non-magnetic CP stars \citep{2007A&A...476..911F, 2008A&A...483..891F, 2009ARA&A..47..333D}. However, a limited number of studies have reported on the detection of a magnetic field in \dSct\ and Am stars \citep[e.g.][]{2008MNRAS.386.1750K, 2015MNRAS.454L..86N, 2023MNRAS.526L..83H}.

Axial rotation is an important factor governing stellar pulsations, but it is not straightforward to fully take it into account when analysing observations and calculating theoretical models.
In the case of a slowly rotating pulsating star, the axial rotation already causes a shift in the observed frequencies \citep{1949MSRSL...9....3L, 1977ApJ...217..151H}. 
This scenario becomes more complicated when fast rotation also alters the shape of the star into an oblate spheroid, as it changes the stellar interior and potentially influences other physical processes that contribute to the pulsation mechanism. 
Convection is one of the physical processes that also affect pulsation. When convective core overshooting occurs, the core boundary is extended, leading to extra mixing in the stellar interior and enhancing the main-sequence lifetime of a star as the core gets more fuel.
There are very few studies about the impact of rotation on the corresponding overshooting parameter $\alpha_{\rm ov}$. 
Therefore, in-depth asteroseismic studies of rapidly rotating stars are required to test and improve our current understanding of the driving of pulsations.
In contrast to the magnetic roAp stars, their non-magnetic pulsating counterparts usually show much weaker modulation of the brightness, which is observed in the most precise space-based light curves and can be attributed to the axial rotation \citep[e.g.,][]{2011MNRAS.415.1691B, 2019MNRAS.490.2112B}. Hybrid \dSct-\gDor\ pulsators and unresolved binaries can show peaks in the same frequency region, where the rotationally-induced modulation is expected \citep[e.g.,][]{2019MNRAS.487.4695S, 2020MNRAS.498.2456S, 2020MNRAS.491.4345R}.
This limits the selection of rapidly rotating stars to measure the broadening of absorption lines in spectra or search for rotationally split frequencies in time-series data. 
The spectroscopic survey of \cite{1997A&AS..122..131S} revealed that \dSct\ stars show a broad distribution in their projected rotational velocity (\vsini) and they found an average value $\langle$\vsini$\rangle$ of 97\,\kms\ for the low-amplitude pulsators and a maximum value exceeding 200\,\kms.

More than two decades ago, the Nainital-Cape Survey was initiated by astronomers from India and South Africa to detect and study the pulsational variability in Ap and Am/Fm stars \citep[e.g;][]{2000BASI...28..251A, 2001AA...371.1048M,2003MNRAS.344..431J,2006A&A...455..303J, 2009A&A...507.1763J, 2010MNRAS.401.1299J, 2012MNRAS.424.2002J, 2016A&A...590A.116J, 2017MNRAS.467..633J}. 
Meanwhile, the project developed into a multi-national collaborative venture for which the most recent results have been published by \citet{2020MNRAS.492.3143T}, \citet{2021MNRAS.504.5528T}, \citet{2022MNRAS.510.5854J}, and \citet{2023MNRAS.524.1044T}.
HD\,118660 is one of the targets of this survey for which photometric observations were gathered that were analysed by \citet{2006A&A...455..303J}.
They discovered \dSct-type pulsations with prominent pulsation periods of $\sim$ 1 and 2.52\, hours.
The subsequent combined spectroscopic and spectropolarimetric study of \cite{2017MNRAS.467..633J} revealed that this star has a sub-solar Ca abundance of $-5.92\pm 0.32$ and shows an excess in Fe of $-4.76\pm 0.23$, where the solar values are $-5.73$ and $-4.53$, respectively, and elements of the lanthanides group typical for the mild metallic-line stars. Given the absence of a detectable magnetic field in HD\,118660 and derived abundances, this object deserves special attention due to exceptionally high \vsini-value of $108 \pm 8$\,\kms close to the upper limit ($\sim$120\,\kms) observed for the Am stars \citep{2009AJ....138...28A}.
Moreover, these authors also observed line profile variations in two consecutive short-exposure spectra with a high signal-to-noise ratio (SNR).
These facts motivated us to perform a comprehensive study of HD\,118660 using modern observational techniques and up-to-date theoretical models.
In the meantime, HD\,118660 has been subjected to a multi-site campaign to acquire time-resolved spectroscopic data. The results based on these data will be published in a forthcoming paper (Semenko et al., in preparation).
A preliminary analysis of the \tess\ time-series data for the two currently available sectors shows a rich frequency spectrum with clear evidence of amplitude modulation due to the beating of two close frequencies \citep{2024arXiv240203403S}.
In this paper, we perform a comprehensive analysis of these \tess\ light curves and theoretical modelling in which we investigate the role of core overshooting in the seismic activity of this star.

The structure of the paper is as follows. 
Sect.\,\ref{observations_data_reduction} describes the observational data and processing. 
A detailed presentation of the asteroseismic modelling and results are given in Sect.\,\ref{asteroseismic_modelling}. In Sect.\,\ref{conclusion}, we end with an overview of the main conclusions drawn from our study and a discussion of their relevance.

\section{Observations and data reduction}
\label{observations_data_reduction}

Transiting Exoplanet Survey Satellite \citep[\tess;][]{2015JATIS...1a4003R} is a NASA space-based observatory launched on 18th April 2018 into a 13.7-day cislunar orbit maintained by a 2:1 lunar resonance. The primary goals of the mission are finding new exoplanets and characterising previously known systems using the transiting method. Moreover, the data collected by \tess\ are ideally suited for detailed asteroseismic studies of pulsating variables. 

HD\,118660 was observed by the $\tess$ in sectors 23 (S23; from March 19, 2020, to April 15, 2020) and 50 (S50; from March 26, 2022, to April 22, 2022).

\begin{figure*}
    \centering
    \includegraphics[width=0.9\textwidth]{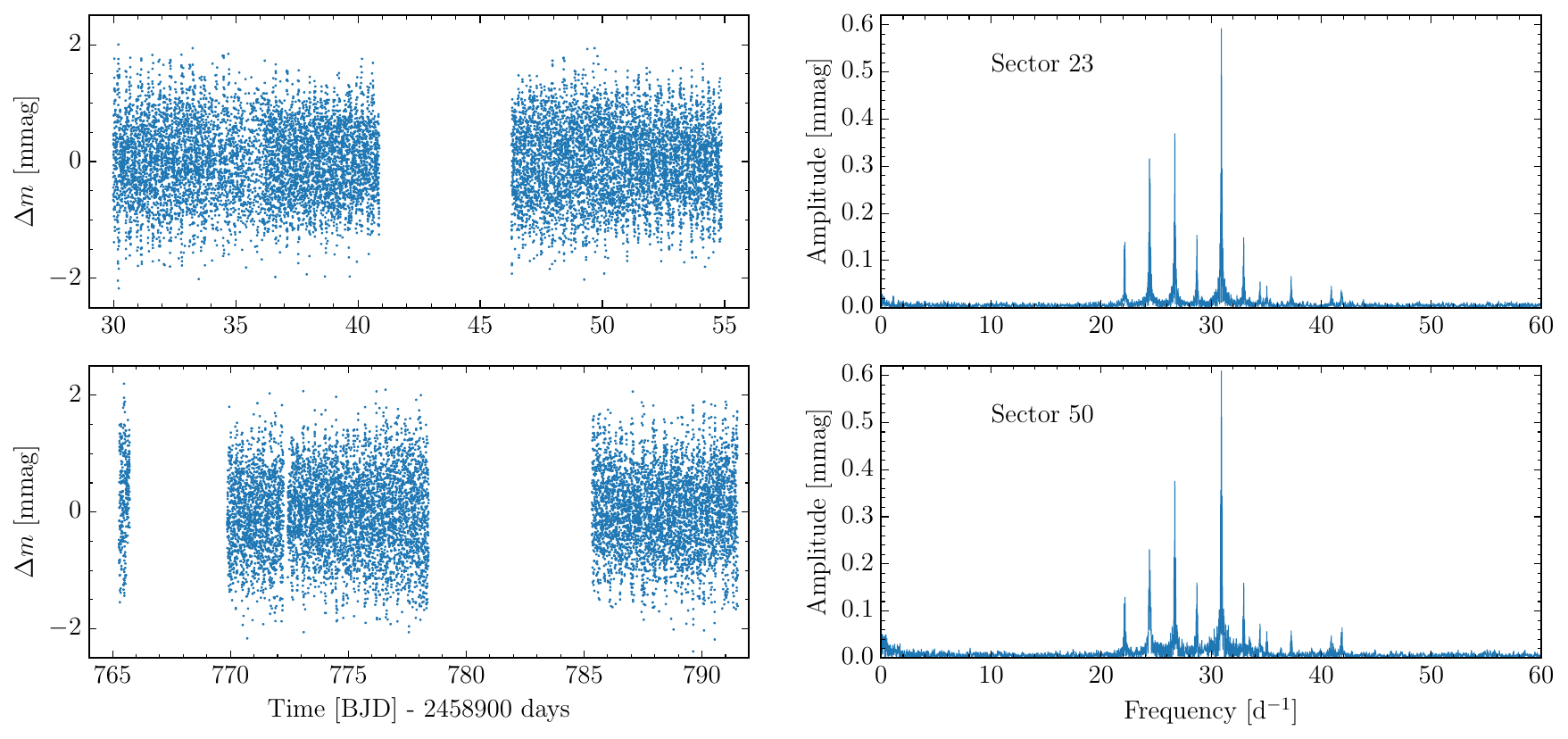}
    \caption{
    Top row: The left panel displays the \tess\ light curve observed in short cadence mode (2-min sampling) in sector 23, and its corresponding Fourier transform is given in the right panel. 
    Bottom row: Idem, but for the data obtained in \tess\ sector 50.
    }
    \label{fig:TimeSeries}
\end{figure*}

We used the Python module \textsc{lightkurve} \citep{2018ascl.soft12013L} to retrieve the \tess\ data of HD\,118660 from the Barbara A. Mikulski Archive for Space Telescopes (MAST) \citep{2019AAS...23324511F}. This object was observed in short cadence mode with a time sampling of 2 min.
The PDCSAP (Pre-Data Conditioned Simple Aperture Photometry) flux data, corrected for long-term trends and systematic errors, were considered for analysis. 
The resulting light curves and the corresponding frequency spectra of HD\,118660 generated using Discrete Fourier Transform (DFT) in \textsc{PERIOD04} \citep{2005CoAst.146...53L} are depicted in the upper and lower panels of Fig.\,\ref{fig:TimeSeries} for sector 23 and sector 50, respectively.

The variations corresponding to the significant frequencies were extracted from the \tess\ light curves. We used the software PERIOD04 for this purpose, which performs DFT of the input time-series data. The significance evaluation of frequency peaks is based on a signal-to-noise ratio (SNR) criterion. As recommended by \citet{2015MNRAS.448L..16B}, only frequencies with an amplitude exceeding five times the mean noise level (SNR$>$5) were retained, where the mean noise level is evaluated near the Nyquist limit of the frequency spectra. 


The uncertainties of the frequencies are calculated as \begin{equation}
    \sigma_{\rm f} = 0.44\,\frac{1}{T_{\rm obs}}\,\frac{1}{\rm SNR},
\end{equation}
where $T_{\rm obs}$ is the time span of the observations \citep[e.g.][]{1999DSSN...13...28M,2012IAUS..285...17K}. 
To determine the errors of the amplitudes and phases, we have performed a 10000-step Monte Carlo test in \textsc{PERIOD04}. 
The resulting significant frequencies detected along with their corresponding amplitudes, phases, and SNR values are tabulated in Table\,\ref{tab:freqID}. 
In our analysis, we considered frequencies larger than $1d^{-1}$ for both S23 and S50 observations. As for S23, there have no significant frequencies below that level and for S50 the noise level is too high.

\begin{table*} 
\centering
    \caption{
    The significant frequencies detected in the variations of the 2-min cadence \tess\ data of sectors 23 (top part) and 50 (bottom part). 
    Each record lists the identifier (ID), the frequency value $f$ and its error $\sigma_{f}$, the degree $\ell$ and order $n$ of the pulsation mode it has been identified with (if available), the amplitude $A$ and its error $\sigma_{A}$, the phase $\phi$ and its error $\sigma_{\phi}$, and the SNR value of the corresponding peak in the DFT.
    An additional ID is given in the second column for the radial modes detected in both sectors (IDrad).
    }
    \label{tab:freqID}
    \begin{tabular}{cccccccccc}
    \hline
     ID & IDrad & $f$ (\cd) & $\sigma_{f}$ (\cd) & $(\ell, n)$ & $A$ (mmag) & $\sigma_{A}$ (mmag) & $\phi$ & $\sigma_{\phi}$ & SNR  \\
    \hline
      \multicolumn{9}{c}{Sector 23} \\
      $f_{23,1}$  & $f_{\rm r,1}$ & 30.9202 & 0.0001 & (0, 5) & 0.5911 & 0.0033 & 0.4141 & 0.0008 & 131\\
      $f_{23,2}$  & $f_{\rm r,2}$ & 26.6894 & 0.0002 & (0, 4) & 0.3691 & 0.0032 & 0.5203 & 0.0013 &  82\\
      $f_{23,3}$  &               & 24.3829 & 0.0002 & -      & 0.2794 & 0.0037 & 0.2817 & 0.0021 &  62\\
      $f_{23,4}$  &               & 24.4431 & 0.0004 & (1, 3) & 0.2053 & 0.0037 & 0.1425 & 0.0028 &  46\\
      $f_{23,5}$  &               & 28.6945 & 0.0005 & (1, 4) & 0.1561 & 0.0032 & 0.1425 & 0.0032 &  35\\
      $f_{23,6}$  &               & 32.9514 & 0.0005 & (1, 5) & 0.1467 & 0.0032 & 0.7954 & 0.0035 &  33\\
      $f_{23,7}$  & $f_{\rm r,3}$ & 22.1567 & 0.0007 & (0, 3) & 0.1101 & 0.0035 & 0.3288 & 0.0051 &  24\\
      $f_{23,8}$  &               & 22.0904 & 0.0010 & -      & 0.0805 & 0.0035 & 0.6042 & 0.0064 &  18\\
      $f_{23,9}$  &               & 37.2650 & 0.0012 & (1, 6) & 0.0666 & 0.0032 & 0.4172 & 0.0077 &  15\\
      $f_{23,10}$ &               & 34.4361 & 0.0014 & -      & 0.0578 & 0.0032 & 0.4813 & 0.0090 &  13\\
      $f_{23,11}$ & $f_{\rm r,4}$ & 35.0489 & 0.0016 & (0, 6) & 0.0479 & 0.0032 & 0.6645 & 0.0107 &  11\\
      $f_{23,12}$ &               & 40.9176 & 0.0018 & -      & 0.0431 & 0.0032 & 0.9283 & 0.0117 &  10\\
      $f_{23,13}$ &               & 41.8077 & 0.0020 & (1, 7) & 0.0394 & 0.0033 & 0.2859 & 0.0134 &   9\\
      $f_{23,14}$ &               & 41.8920 & 0.0020 & -      & 0.0384 & 0.0014 & 0.1522 & 0.3292 &   9\\
      $f_{23,15}$ &               &  1.0846 & 0.0030 & -      & 0.0246 & 0.0121 & 0.4911 & 0.2345 &   6\\
      \hline 
      \multicolumn{9}{c}{Sector 50} \\
      $f_{50,1}$  & $f_{\rm r,1}$ & 30.9204 & 0.0001 & (0, 5) & 0.6121 & 0.0047 & 0.1694 & 0.0012 & 107\\
      $f_{50,2}$  & $f_{\rm r,2}$ & 26.6889 & 0.0002 & (0, 4) & 0.3741 & 0.0047 & 0.7713 & 0.0020 &  65\\
      $f_{50,3}$  &               & 24.3792 & 0.0002 & -      & 0.2806 & 0.0057 & 0.8311 & 0.0033 &  49\\
      $f_{50,4}$  &               & 24.4383 & 0.0004 & (1, 3) & 0.2120 & 0.0058 & 0.2648 & 0.0041 &  33\\
      $f_{50,5}$  &               & 28.6937 & 0.0005 & (1, 4) & 0.1586 & 0.0048 & 0.8905 & 0.0053 &  35\\
      $f_{50,6}$  &               & 32.9518 & 0.0006 & (1, 5) & 0.1553 & 0.0047 & 0.2479 & 0.0048 &  27\\
      $f_{50,7}$  & $f_{\rm r,3}$ & 22.1629 & 0.0007 & (0, 3) & 0.1048 & 0.0057 & 0.9433 & 0.0085 &  24\\
      $f_{50,8}$  &               & 34.4345 & 0.0010 & -      & 0.0664 & 0.0049 & 0.9410 & 0.0115 &  18\\
      $f_{50,9}$  &               & 41.8914 & 0.0012 & -      & 0.0669 & 0.0058 & 0.8558 & 0.0264 &  15\\
      $f_{50,10}$ &               & 22.0905 & 0.0013 & -      & 0.0858 & 0.0057 & 0.7898 & 0.0102 &  13\\
      $f_{50,11}$ &               & 37.2663 & 0.0016 & (1, 6) & 0.0573 & 0.0047 & 0.8712 & 0.0131 &  11\\
      $f_{50,12}$ & $f_{\rm r,4}$ & 35.0481 & 0.0018 & (0, 6) & 0.1169 & 0.0099 & 0.4763 & 0.0296 &  10\\
      $f_{50,13}$ &               & 40.9137 & 0.0020 & -      & 0.0491 & 0.0006 & 0.3884 & 0.0778 &   9\\
      $f_{50,14}$ &               & 33.4568 & 0.0020 & -      & 0.0438 & 0.0006 & 0.4374 & 0.0425 &   9\\
      $f_{50,15}$ &               & 29.1510 & 0.0025 & -      & 0.0312 & 0.0006 & 0.1043 & 0.1100 &   7\\ 
      $f_{50,16}$ &               &  1.0983 & 0.0035 & -      & 0.0239 & 0.0118 & 0.1981 & 0.1160 &   5\\
      \hline
\end{tabular}
\end{table*}

\section{Asteroseismic modelling}
\label{asteroseismic_modelling}

\subsection{Modelling Parameters}
In our analysis, we performed model-based seismic analysis, estimating the seismic parameters followed by mode identification of the pulsation frequencies.
The stellar evolution models were created using the Code Li\'{e}geois d'Evolution Stellaire (CL\'{E}S; \citealt{2008Ap&SS.316...83S}). The step-wise radial eigenfrequencies were calculated using the Li\`{e}ge Oscillation Code (OSC; \citealt{2008Ap&SS.316..149S}). 
The grid of models encompasses a mass range of 1.10--2.00\,$M_{\odot}$ with a step size of 0.05\,$M_{\odot}$.
The metallicity ($Z$) varies from 0.002 to 0.020 with a step of 0.002. 
Finally, the grid incorporates a convective overshooting parameter ($\alpha_{\rm ov}$) values of 0.1, 0.2 and 0.3. 
The stellar model calculation was started with an initial H-abundance of $75\%$ assuming that the star is on the zero-age main sequence (ZAMS) and continued until the star reaches the red giant branch. 
The estimated rotation velocity of the star is nearly 28\% of the critical value. For \dSct\ stars a 2D numerical approach shows that that rotation velocity below the 50\% of the critical value, the low-order modes ($\ell=0,1$) normalized by the large separation, these modes remain fairly constant \citep{10.3389.fspas.2022.934579}. Therefore, in our model calculations, we have not considered the effect of rotation.



\subsection{Mode Identification} 
\label{sec: mode identification}
The \dSct\ variables pulsate in low-order overtones of $p$ modes with pulsation frequencies above 5\,d$^{-1}$. 
Two techniques were used to identify the modes: the interpretation of the \'{e}chelle diagram and the analysis of frequency ratios. 
The latter is also referred to as the theoretical Petersen diagram \citep{1973A&A....27...89P,1978A&A....62..205P}. The following subsections provide the details of the applied procedures.

\subsubsection{\'{E}chelle diagram}
\label{sect:echelle}
In the asymptotic regime, where $n\gg l$, the modes of a given $\ell$ are approximately equally spaced in the frequency domain by the value of parameter $\Delta\nu$, which is known as large frequency separation. The inverse value of $\Delta\nu$ is the time taken by a sound wave to traverse the star. 
The frequencies $\nu$ are related to the order $n$ and degree $\ell$ of the pulsation modes and to the large frequency separation $\Delta\nu$ through the asymptotic equation of the form \citep{1980ApJS...43..469T}
\begin{equation}
        \nu_{n,\ell}=\Delta\nu\,(n+\ell/2+\epsilon),
        \label{asymptotic}
\end{equation}
where $\epsilon$ is a constant of order unity.
It is also well established that the oscillation frequencies (and therefore the large separation $\Delta\nu$) scale as the square root of the mean stellar density ($\bar{\rho}$) by the empirical relation $\Delta\nu\propto \sqrt{\bar{\rho}}$ \citep{2010aste.book.....A}.

Eqn. \,\ref{asymptotic} demonstrates that the radial and non-radial modes appear repetitively with the same overtones for lower $\ell$ in the Fourier transform. 
However, the \dSct\ stars do not pulsate in the asymptotic regime \citep{2020Natur.581..147B}. Furthermore, HD\,118660 is a fast-rotating star and this fact seriously restricts the use of Eqn.\,\ref{asymptotic} for detailed seismic analysis. In our study, we refer to Eqn. \,\ref{asymptotic} only during the identification of radial modes and for the first estimation of the large frequency separation.
 
\begin{figure*}
    \centering
    \includegraphics[width=0.65\columnwidth]{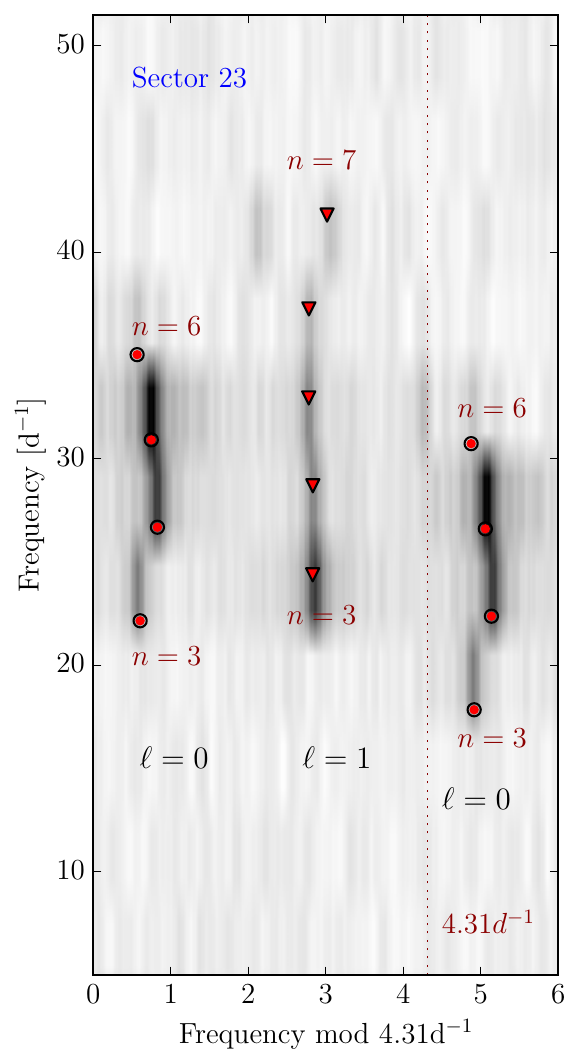}
    \hspace{2cm}
    \includegraphics[width=0.65\columnwidth]{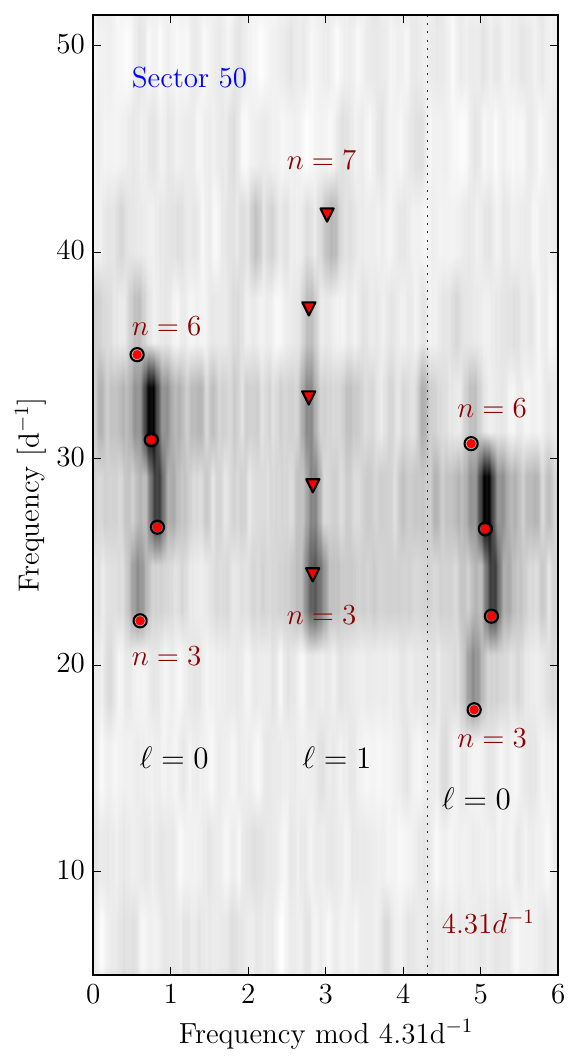}
    \caption{\textit{Left panel:} The \'{e}chelle diagram depicts frequencies derived from the \tess\ light curve in S23. The red circular points represent radial ($\ell=0$) overtones, and the red triangular points are dipolar overtones ($\ell=1$).
    \textit{Right panel}: \'{e}chelle diagram with same observed frequencies for S50.}
    
    \label{fig:ed_model}
\end{figure*}

The \'{e}chelle diagram is an important tool for analyzing spectra of stellar pulsations. 
In this diagram, the frequency spectrum is sliced into equal chunks with a width of $\Delta\nu$ and stacked vertically towards higher frequencies \citep{2020Natur.581..147B}.
We used a dynamic \'{e}chelle interface \citep{2022ascl.soft07005H} to fine-tune the value of $\Delta\nu$. 
In Fig.\,\ref{fig:ed_model}, we show  frequency spectra of HD\,118660 in \'{e}chelle format. 
The best visually aligned pattern with two vertical ridges corresponding to the angular degrees $\ell=0$ (left) and 1 (right) is achieved for $\Delta\nu=4.31$\,d$^{-1}$. 
 
\begin{figure}
    \centering
    \includegraphics[width=\columnwidth]{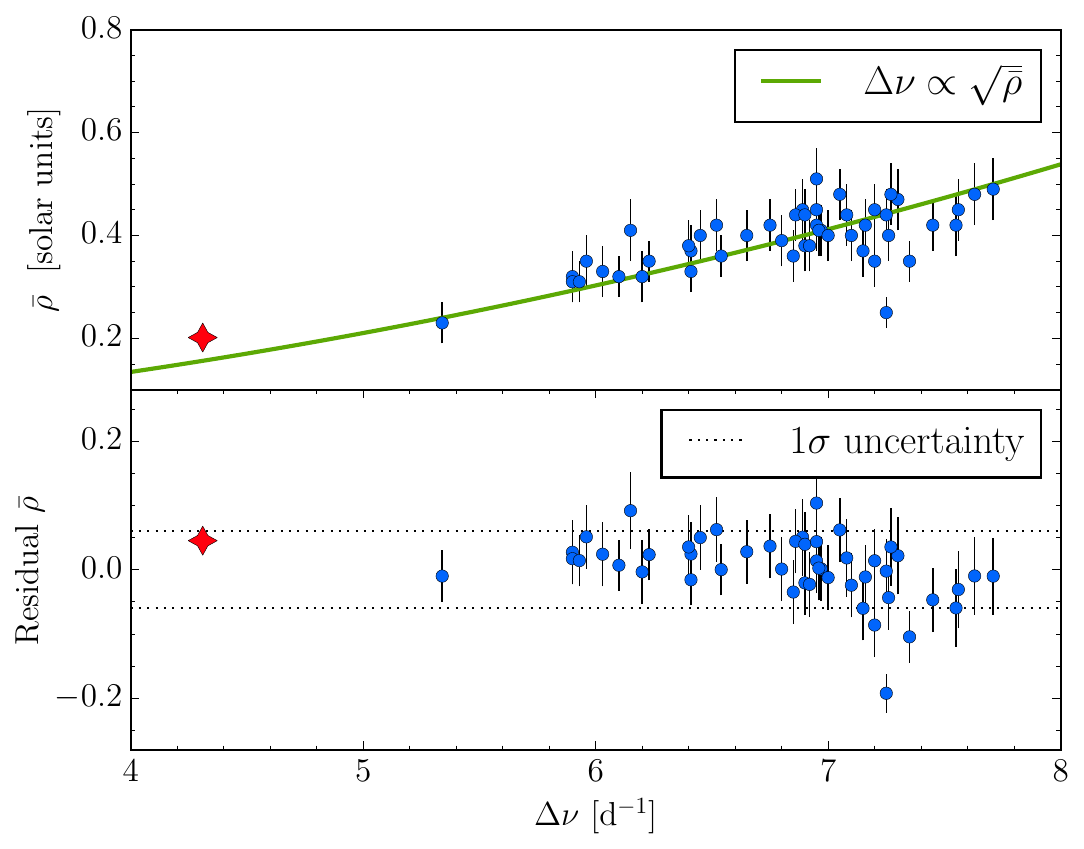}
    \caption{A comparison of HD\,118660 (red asterisk) to a group of \dSct\ stars (blue circles) listed in \citet{2020Natur.581..147B} that indicates that the large separation $\Delta\nu=4.31$\,d$^{-1}$ found in our study for HD\,118660 follows the $\Delta\nu$ vs {$\bar{\rho}$} relation within the $1\sigma$ spread. The error in $\bar{\rho}$ corresponds to the $1\sigma$ spread of the mean density of best-fit models shown in Fig.\,\ref{fig:seismic_plot}.} 
    \label{fig:lar_sep_den}
\end{figure}

For the frequencies detected in the \tess\ data of both sectors that are situated on the radial ridge of the \'{e}chelle diagram, we have introduced an additional identifier in the second column of Table\,\ref{tab:freqID}. 
The frequencies aligned in the vertical direction correspond to the consecutive overtones. 
The estimated value of $\Delta\nu$ and the mean stellar density computed from the best-fit model for HD\,118660 agree well with the corresponding characteristics published in \citet{2020Natur.581..147B} for a group of \dSct\ stars, as shown in Fig.\,\ref{fig:lar_sep_den}.

\subsubsection{Frequency ratios}
\label{sect:ratios}

\begin{figure*}
    \centering
    \includegraphics[width=0.95\textwidth]{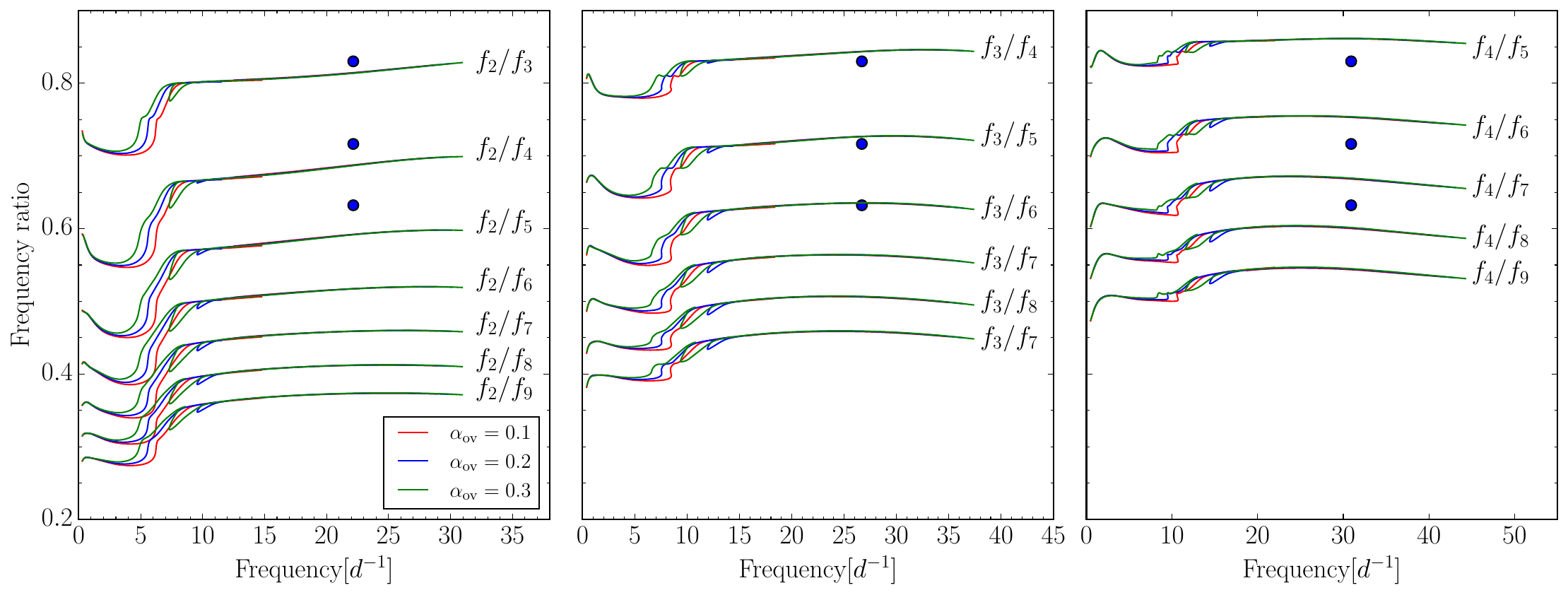}
    \caption{The evolution of frequency ratios for radial modes ($\ell=0$) of consecutive overtones, plotted using theoretical models with $Z=0.014$ and $\alpha_{\rm ov}=0.1,0.2,0.3 $ (in \textit{red, blue} and \textit{green}, respectively). The frequencies in the numerator correspond to the first overtone $f_2$ ($n=2$; left panel), second overtone $f_3$ ($n=3$; middle panel), and third overtone $f_4$ ($n=4$; right panel). The dots denote the frequency ratios of the observed radial modes with $f_{\rm r,3}$ in the numerator.}
    
    \label{fig:peterson}
\end{figure*}

For the frequency ratio analysis, we only considered the observed frequencies associated with radial modes ($\ell=0$), being $f_{\rm r,1}$, $f_{\rm r,2}$, $f_{\rm r,3}$, and $f_{\rm r,4}$ (second column of \,Table\,\ref{tab:freqID}). 
We calculated the frequency ratios for different pairs of radial orders under the assumption that the star is located on the main sequence.
In Fig.\,\ref{fig:peterson}, we show theoretical Peterson diagrams obtained for $Z=0.014$ and $\alpha_{ov}=0.1,0.2,0.3$  (in \textit{red, blue} and \textit{green} colors, respectively).
We use the notation $f_{\rm n+1}$ to refer to the theoretical frequency of the $\it{n^{th}}$ overtone radial mode.
Each panel of Fig.\,\ref{fig:peterson} shows the results for another radial mode in the numerator of the frequency ratios: the first overtone in the left panel ($n=2$), the second overtone in the middle panel ($n=3$), and the third overtone in the right panel ($n=4$). 
The dots represent the frequency ratios of the observed radial modes, being $f_{\rm r,3} / f_{\rm r,2}\sim 0.8303$\,\cd\ (upper dot), $f_{\rm r,3} / f_{\rm r,1} \sim 0.7167$\,\cd\ (middle dot) and $f_{\rm r,3} / f_{\rm r,4} \sim 0.6323$\,\cd\ (lower dot). 
The best visual match is found in the middle panel of Fig.\,\ref{fig:ed_model}, corresponding to the second overtone ($n=3$) in the numerator.

We did the same exercise for the other values of the overshooting parameter $\alpha_{ov}$. It causes small variations of the theoretical frequency ratios for frequencies below 15\,d$^{-1}$ while the frequency ratio values in the flat part of the diagram remain unchanged. 
As the detected radial mode with the lowest frequency is located in the flat region of the Peterson diagrams, we conclude that the identification of the radial overtones does not depend on the value of $\alpha_{ov}$.

Overtones of orders outside the plotted range are not included intentionally. 
The possibility of finding another correlation with the data remains, but we leave it unexplored as, in this case, the large separation must be an integer multiple or fraction of 4.31\,\cd.

\bigskip

The overall conclusion of our efforts to identify the modes (Sections\,\ref{sect:echelle} and \ref{sect:ratios}) is that the radial overtones of orders $n=3$--$6$ are present in the observed variations of the \tess\ data. 
This interpretation was added in Fig.\,\ref{fig:ed_model}.
These radial overtones are almost equidistant in frequency.

\subsection{Stellar Models}
\label{sec:asteroseismic_paramters}

Our study is based upon a grid of 570 evolution tracks calculated for different values of the independent parameters $M$ (19 values), $Z$ (10 values), and $\alpha_{\rm ov}$ (3 values).
We used the seismic-$\chi^{2}$ distribution from \citet{2021MNRAS.502.1633M} to find minima of the $\chi^{2}$ values from each of the grid mentioned earlier.

\begin{figure*} 
    \centering
    \includegraphics[width=0.33\textwidth]{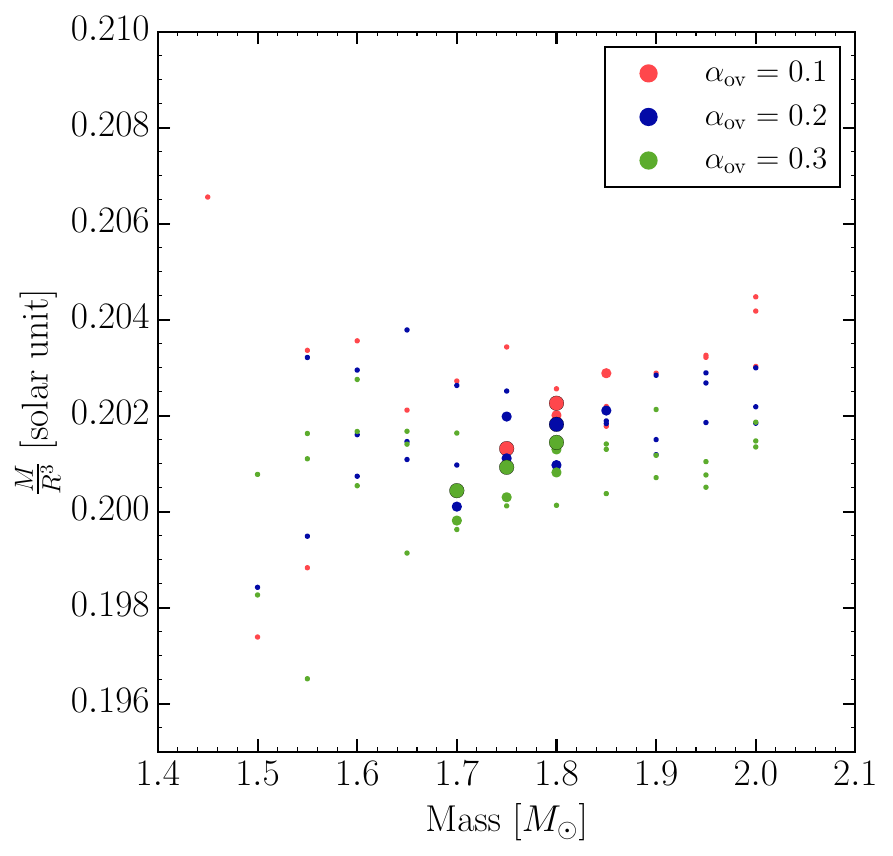}
    \includegraphics[width=0.33\textwidth]{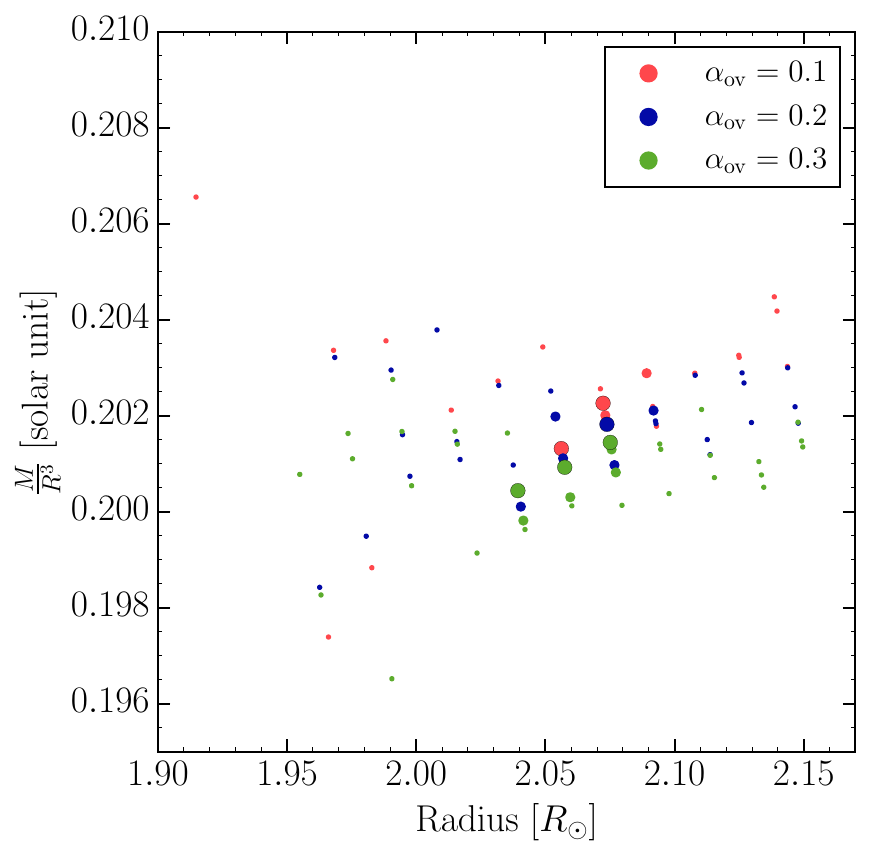}
    \includegraphics[width=0.33\textwidth]{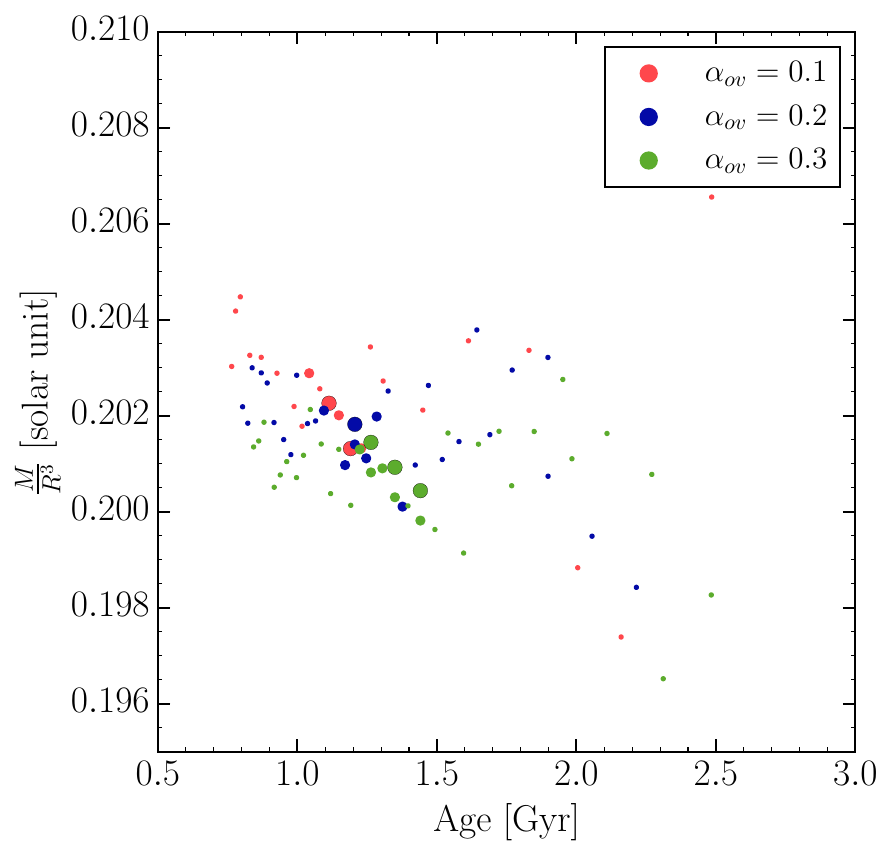}
    \caption{
    Illustration of the relationship between $M/R^3$ and the parameter mass (left panel), radius (middle panel), and age (right panel). 
    The red, blue, and green points represent the models limited by the criteria described in Sect.\,\ref{sec:asteroseismic_paramters} for $\alpha_{ov}=$\,0.1, 0.2, 0.3, respectively. 
    The smallest dots represent all stellar models, while the medium and big sizes indicate models limited within $3\sigma$ and $1\sigma$ from the observed position in the H-R diagram, respectively.
    }
    \label{fig:seismic_plot}
\end{figure*}

In Fig.\,\ref{fig:seismic_plot}, the results of the seismic modelling are shown as the scattered plots of the mean stellar density $\rm M/R^{3}$ versus mass $M$, radius $R$, and age $t$. 
The stellar models, corresponding to the convective overshooting parameters $\alpha_{\rm ov}=0.1, 0.2, 0.3$, are plotted in red, blue, and green, respectively. 
The models constrained within $3\sigma$ and $1\sigma$ from the observed position in the H-R diagram are marked using the medium and big markers. In the following subsections, we outline the details of the modelling.

\subsubsection{Overshooting parameter}
\label{sec:ov}

The overshooting parameter $\alpha_{\rm ov}$ strongly depends upon the mass for the low and intermediate-mass stars \citep{1995MNRAS.276.1287U, 2016A&A...592A..15C,2017ApJ...849...18C, 2018ApJ...859..100C,2019ApJ...876..134C}.  
The value of $\alpha_{\rm ov}$ increases almost linearly for stars in the mass range $M\sim$1.2--2.0\,M$_{\odot}$ and saturates near $\alpha_{\rm ov}\sim0.22$ for stars with $M$ up to 4.4\,M$_{\odot}$. 
Convective overshooting plays a vital role in stellar evolution as it extends the lifetime of stars on the main sequence. 
The significance of overshooting increases with age and mass while the exact value of $\alpha_{\rm ov}$ remains not well-established \citep{2017ApJ...849...38L, 2018ApJ...859..100C, 2021A&A...655A..29J, 2022ApJ...924..130D}. 
Indeed, the uncertainty of $\alpha_{\rm ov}$ at the saturation point can reach up to 0.4 \citep{2018A&A...618A.177C}. 
In light of these observations and considering the spectral type of HD\,118660, our choice to constrain $\alpha_{\rm ov}$ between 0.1 -- 0.3 for the seismic analysis appears to be well justified.

The overshooting parameter $\alpha_{\rm ov}$ was implemented in the stellar evolution code CL\'{E}S to calculate the size of the overshooting zone $r_{\rm ov}$, given by 
\begin{equation}
    r_{\rm ov} = r_{c} \pm \alpha_{\rm ov} \min(H_{P},h),
\end{equation}
where $r_{c}$ is the radius of the convective zone at the boundary, $h$ is its size, and $H_{\rm P}$ is pressure scale height.

\begin{figure*} 
    \centering
    \includegraphics[width=0.46\textwidth]{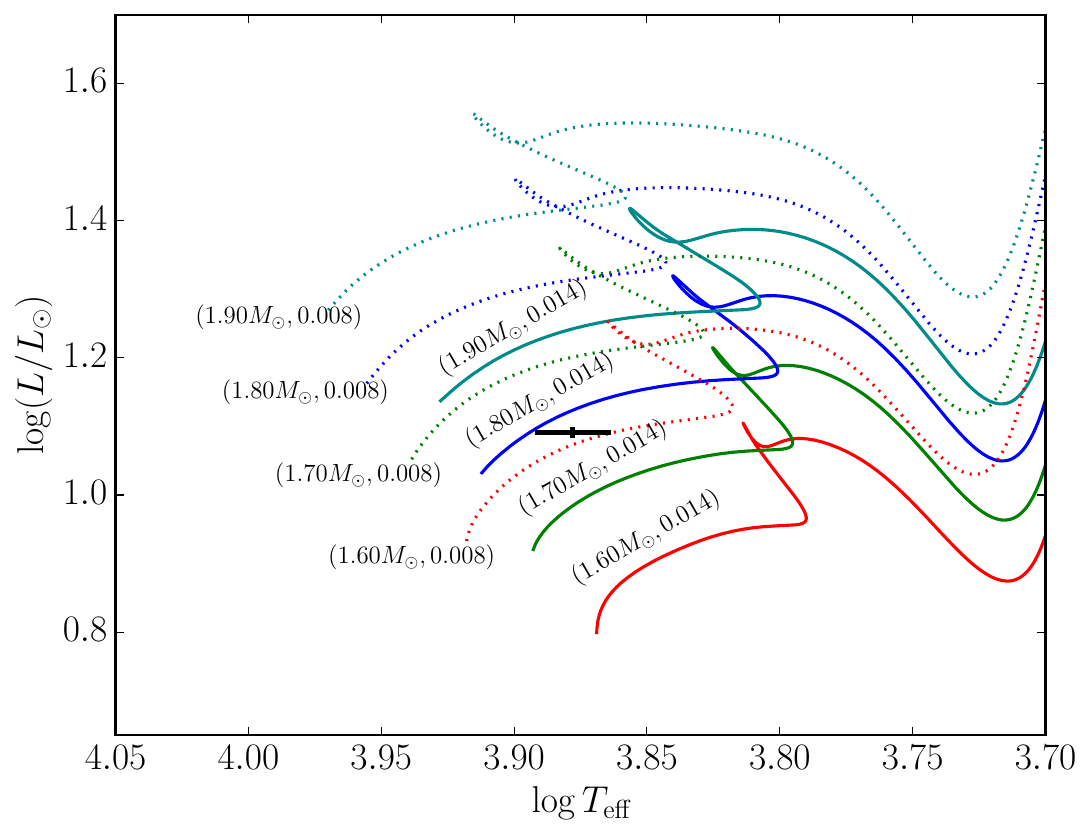}
    \includegraphics[width=0.46\textwidth]{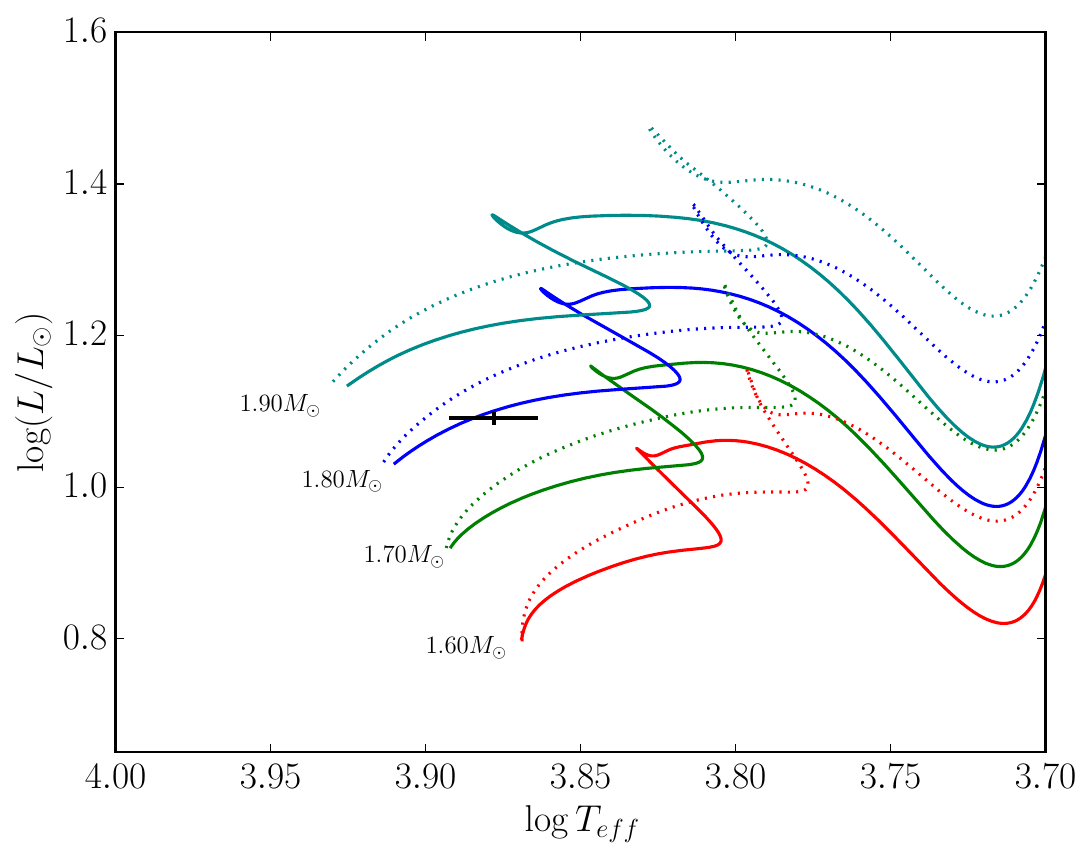}
    \caption{
    Left panel: The evolution tracks corresponding to the global metallicities $Z=0.008$ (dotted lines) and $Z=0.014$ (solid lines) computed for $\alpha_{\rm ov}=0.2$. 
    Right panel: The evolution tracks calculated for $\alpha_{\rm ov}=0.1$ (solid lines) and $\alpha_{\rm ov}=0.3$ (dotted lines) for solar metallicity ($Z=0.014$).
    The black error bar represents the position of HD\,118660 on the H-R diagram.}
    \label{fig:param_var}
\end{figure*}

To demonstrate the importance of $\alpha_{\rm ov}$ as a seismic parameter, we show two different sets of evolution tracks computed for $\alpha_{\rm ov}=0.1$ (solid line) and $0.3$ (dotted line) for $Z$ fixed at $0.014$ in the right panel of Fig.\,\ref{fig:param_var}. The position of HD\,118660 is shown with black mark.
One can notice that the overshooting parameter has an increasing effect on the seismic age in the main-sequence phase. 
With higher values of $\alpha_{\rm ov}$, the more intensive conveyance of fresh hydrogen in the stellar core extends the stellar lifespan. As a result, the seismic age increases with the overshooting parameter (Table \ref{tab:sigma}). However, the impact of $\alpha_{\rm ov}$ on the rest of seismically estimated stellar parameters is less significant. 
For example, our study shows that the seismic parameters of HD\,118660, are independent of $\alpha_{\rm ov}$ except for the stellar age.

\begin{figure}
    \centering
    \includegraphics[width=0.9\columnwidth]{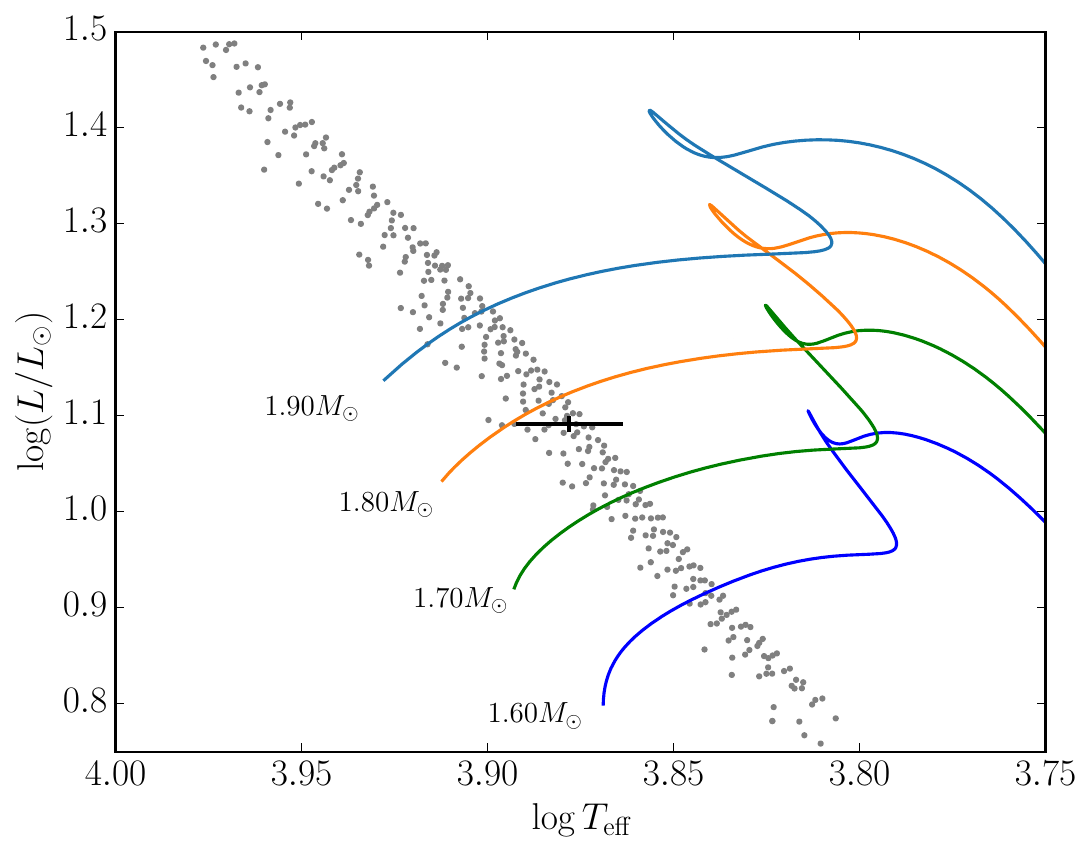}
    \caption{
    The grey dots show the models with the lowest $\chi^{2}$ value from each of the 570 evolution tracks mentioned in Section\,\ref{sec:asteroseismic_paramters} 
    overlaid with evolution tracks for $Z=0.014$ and $\alpha_{\rm ov}=0.2$. 
    The black error bars show the location of HD\,118660 in the H-R diagram.
    } 
    \label{fig:hr_constrained}
\end{figure} 

\subsubsection{Metallicity}
\label{sec:meta}

The stellar metallicity $Z$ is also an important parameter used in seismic modelling, directly affecting the seismic age. 
%
A higher value of $Z$ results in a higher seismic mass $M$. 
Consequently, the seismic age $t$ of the star appears to be lower, illustration can be seen in  ($\alpha_{\rm ov}=0.2$) in the left panel of Fig.\,\ref{fig:param_var}. The tracks computed for $Z=0.008$ and $Z=0.014$ are given with dotted and solid lines, respectively. 
%

It demonstrates that an increase of $Z$ results in a shift of the evolutionary tracks towards the lower values of both the luminosity $L$ and effective temperature \teff, resulting in a higher value of the seismic mass $M$.


Atomic diffusion is responsible for the emergence of irregularities in the horizontal and vertical distribution of chemical elements in stellar atmospheres. It distorts the observed metallicity and hence complicates asteroseismic analyses. 
Membership of an open cluster can potentially resolve the global metallicity problem, but HD\,118660 is a field star located in the solar vicinity at a distance of $\sim70$\,pc (calculated from the \textit{Gaia} DR3 parallax). 
Along with that, the estimated iron abundance of the star is close to the solar value \citep{2017MNRAS.467..633J}.
In this analysis, we constrain the global metallicity ($Z$) value close to solar ($Z$). 
We performed the modelling by varying $Z$ within $0.002$ of the solar value \citep[$Z_{\odot}=0.014$;][]{2021A&A...653A.141A,2024A&A...681A..57B}. 
Irrespective of the exact value of $Z$, HD\,118660 is situated within the main-sequence band of the H-R diagram. 

\subsubsection{Age, luminosity, and effective temperature}
\label{sec:ep}

\citet{2017MNRAS.467..633J} showed that HD\,118660 is a main sequence star.
We used this fact to restrict our grid of models within the main sequence region of the H-R diagram.

%
%
%
%
The standard luminosity relation is used to calculate an updated value of the luminosity \luminosity\ of HD\,118660.
Indeed, a more accurate value of $\pi=14.2029\pm0.0275$\,mas for the parallax of HD\,118660 has been published recently in the third data release of {\it Gaia} \citep{2023A&A...674A...1G}. 
For the interstellar extinction, we used $A_\mathrm{V}=0$\,mag. It was derived from the interstellar dust map catalogued by \citet{2019ApJ...887...93G}.
Like \citet{2017MNRAS.467..633J}, the bolometric correction $BC=0.033$ was taken from \citet{1996ApJ...469..355F} and the bolometric magnitude of the sun $M_{\rm bol,\odot}=4.74$ was published by \citet{Willmer_2018}, $m_{v}$ is taken as $6.44$ \citep{2020yCat.1350....0G}, which is similar to the mean photometric magnitude observed in TESS band in both the sectors.
In this way, we derived \luminosity$ =1.09\pm0.01$ $ \rm dex$  for HD\,118660. This value is slightly lower than $1.12\pm0.27$ published by \citet{2017MNRAS.467..633J}, but they are compatible if the errors are considered.
%
We adopted the effective temperature $T_{\rm eff}=7550\pm150$\,K measured by \citet{2017MNRAS.467..633J} in combination with our updated value of the luminosity \luminosity$=1.09\pm0.01$ $\rm dex$ to define the observed location of HD\,118660 in the H-R diagram in Figs.\,\ref{fig:param_var} and \ref{fig:hr_constrained}.

One can notice from Fig.\,\ref{fig:hr_constrained} that the application of the seismic-$\chi^{2}$ criterion results in stellar models scattered over a large range of $T_{\rm eff}$ and $\log L/L_{\odot}$. 
The models at a large distance for the observed positon of HD\,118660 can be safely ignored. 
At the same time, the strict seismic constraints arising from the characteristics of the identified radial overtones define a dense band that includes the observed position of HD\,118660 in the H-R diagram.

\begin{table*} 
\centering
    \caption{The final values of metallicity, mass, radius, and age calculated in the current study using the mean values of models constrained within $1\sigma$ error of the observed $\log (L/L_{\odot})$ and $\log T_{\rm eff}$. And the error corresponding to the mass, age and $v_{\rm eq}$ mentioned here for stellar models with $\alpha_{ov}$  showing largest spread.} 
    \label{tab:sigma}
    \begin{tabular}{cccccc}
    \hline
     $\alpha_{\rm ov}$ & $\frac{M}{M_{\odot}}$ & $\frac{R}{R_{\odot}}$ & Age ($\rm Gyr$) & $\frac{M}{R^{3}}$ [solar]  & $\upsilon_{\rm eq}$ (\kms) \\
     \hline
     0.1 &  $1.77\pm0.04$ & $2.06\pm0.01$ & $1.15\pm0.07$ & 0.20& $114\pm1$  \\
     0.2 & $1.80\pm0.04$ & $2.07\pm0.01$ & $1.21\pm0.07$  & 0.20& $115\pm1$\\
     0.3 &  $1.75\pm0.04$ & $2.06\pm0.01$ & $1.35\pm0.07$ & 0.20& $114\pm1$\\
    \hline
    \end{tabular}
\end{table*}

\subsubsection{Stellar Rotation}
\label{stellar_rotation}
For the current study, we ignored the effects of stellar rotation in our grid of models. 
Most CP stars are found to be slow rotators \citep{1965ApJ...142.1594C, 1995ApJS...99..135A}. However, a recent study showed that the projected rotational velocity \vsini\ for Am stars can reach up to $120$\,\kms\ and shows that they are younger compared to non-CP stars with the same rotation rate \citep{2021AJ....162...32Q}.
For HD\,118660, the available high-resolution spectra have revealed that it is a relatively fast rotator with a \vsini\ value of about $100$--$110$\,\kms\ \citep{2014MNRAS.441.1669C, 2017MNRAS.467..633J, 2023A&A...674A...1G}. 
Depending on the value of the inclination angle $i$, the rotation velocity of HD\,118660 can be high enough to make our simplified approach unacceptable. 
To explore this problem in detail, we have analysed the available photometric data for any signature of stellar rotation.


The Keplerian critical (breakup) rotational velocity \citep{2019A&A...625A..88G} can be calculated as 
\begin{equation} \label{eq:vc}
\upsilon_\mathrm{c}=\sqrt{G\,M/R},
\end{equation}
where $G$ is the gravitational constant. 
By using the values $R=2.07\,\rm R_\odot$ and $M=1.8\,\rm M_\odot$ for the equatorial radius and mass, respectively (second line of Table\,\ref{tab:sigma}), we find $\upsilon_\mathrm{c}=407$\,\kms\ for HD\,118660.

The equatorial rotational velocity $\upsilon_\mathrm{eq}$ is given by 
\begin{equation}\label{eq:ve}
    \upsilon_{\rm eq}=50.6\,R\,f_{\rm rot},
\end{equation}
By using $R=2.07\,\rm R_\odot$ (see Table\,\ref{tab:sigma}) and by considering $\upsilon_\mathrm{c}=407$\,\kms\ as the maximum and \vsini\,$=108$\,\kms\ from \citet{2017MNRAS.467..633J} with $i=90^{\circ}$ as the minimum value for $v_{\rm eq}$, respectively, we can estimate the expected range for the rotational frequency $f_{\rm rot}$ of HD\,118660 to be $\approx 0.9$--$4$\,\cd.

In the frequency spectra calculated for the \tess\ light curves, we find only two peaks with frequencies falling in the specified range: $f_{23,15}\approx1.08$\,\cd\ for S23 and $f_{50,16}\approx1.09$\,\cd\ for S50 (Table\,\ref{tab:freqID}). 
If we interpret the average value $1.085\pm0.003$\,\cd\ (error is corresponding to the $\sigma_{f}$ of $f_{23,15}$ and $f_{50,16}$  from Table \ref{tab:freqID}) as $f_{\rm rot}$, then we can use Eq.\,\ref{eq:ve} to estimate the angle $i$ between the rotational axis and the line-of-sight. 
In this way, we derived that $\upsilon_\mathrm{eq}\approx114$\,\kms\ and $i\approx71^{\circ}\pm2^{\circ}$ (\vsini\ 
 taken from \citet{2017MNRAS.467..633J}.
Hence, we find that $\upsilon_{eq}/\upsilon_{c}\approx0.28$. 
As this value is below the limit of $\upsilon_{eq}/\upsilon_{c}=0.4$ that is set as the limit for the rotational effects to become significant \citep[e.g.,][]{2012A&A...537A.146E}, this would imply that we can consider our modelling of the stellar pulsations of HD\,118660 as adequate.

\section{Conclusions}
\label{conclusion}

This paper presents a detailed seismic study of HD\,118660, a \dSct\ pulsator with mild anomalies of its surface chemical composition. 
Our interest in this particular star is based on its brightness and the accessibility of high-quality and time-resolved photometric and spectroscopic observations. 
\citet{2014MNRAS.441.1669C} and \citet{2017MNRAS.467..633J} classify this star as a fast rotator with a \vsini-value around 100\,km\,s$^{-1}$, which is close to the observed upper limit for Am stars. 
Rapid axial rotation remains challenging to account for in asteroseismic modelling and new studies may help to verify the progress achieved in theory. 
In our study, the effects of rotation are not incorporated.  
Instead, we are focused on the seismic determination of the mass, radius, and age of HD\,118660, as well as on the study of the effects of convective overshooting.

This part of the study is based solely on the currently available \tess\ observations from sectors 23 and 50 and the interpretation and modelling of the observed variations. 
We estimated the seismic parameters after identifying the pulsation modes detected in the \tess\ photometry. 
To identify the modes, we used an \'{e}chelle diagram and compared the frequency ratios in the theoretical Petersen diagram. 
This allowed us to conclude that the radial overtones with orders $n=3-6$ are present in the data of HD\,118660. 
Subsequent tests based on the seismic-$\chi^{2}$ distribution have been used to find the best-fit model from an appropriate grid of theoretical models calculated with the code \textsc{cl\'{e}s}.

HD\,118660 is a late A-type \dSct\ pulsator with $T_{\rm eff}=7550\pm150$\,K \citep{2017MNRAS.467..633J} located near to the red edge of the \dSct\ instability strip \citep{2004A&A...414L..17D}. 
Theory predicts the excitation of high overtones near the blue edge of the instability strip. 
The observed pulsation pattern of HD\,118660 does not seem consistent with these predictions for stars with a solar metallicity. 
As the global metallicity of HD\,118660 hardly differs from the one of the Sun, one can expect that the process of the chemical transport, which is responsible for the chemical peculiarity, modifies the opacity and thus the $\kappa$-mechanism in such stars. 
We intend to address this problem in a follow-up study.

In the current study, we found that the constraints of the radial overtones are not sufficient to constrain the overshooting parameter ($\alpha_{ov}$) and metallicity ($Z$). Therefore they are subjected to additional constraints to estimate the seismic parameter.
Fig.\,\ref{fig:hr_constrained} demonstrates that the known radial overtones can only produce a cloud of luminosities and effective temperatures spread over a huge range. 
For the final solution, we have selected those stellar models having $\log(L/L_{\odot})$ and $\log T_{\rm eff}$ within $1\sigma$ from the values derived from observations. 
The problem with an unconstrained metallicity appears even more critical. 
In this case, we have adopted $Z=0.014\pm0.002$, based on an {\it a priori} knowledge of the proximity of the studied star to the Sun. 
With the metallicity being fixed, the estimated seismic age of HD\,118660 ranges within $1.15$--$1.35$\,Gyr that manifests the star HD 118660 is a ZAMS phase of evolution.

At the same time, we cannot exclude that the real value of the global metallicity of HD\,118660 differs slightly (but enough to impact the modelling), depending on the specific origin. 
We have not elaborated further on this problem as it extends far beyond the scope of the present study. 
We conclude that our study has shown that with the identified radial modes only, asteroseismology cannot provide sufficient constraints on the convective overshooting and metallicity in \dSct\ stars.

Finally, the seismic study of HD\,118660 has also confirmed that the star is a fast rotator. 
In the {\it TESS} data, we have identified at least one frequency that can most likely be attributed to rotational modulation. 
This frequency appears in the frequency spectra of \tess\ data obtained in both sectors (S23 and S50). 
If we interpret it as the rotational frequency $f_{\rm rot}$ and use it in combination with the value for the equatorial stellar radius as found in our study, we could estimate the equatorial rotational velocity to be $\sim$114\,km\,s$^{-1}$. 
This value corresponds approximately to { 28\%} of the breakup rotational velocity for an A-type star with the characteristics of HD\,118660. 
We plan to assess the rotational properties of this star in an upcoming paper containing an in-depth spectroscopic study of HD\,118660.

\section*{Data Availability}

TESS time-series flux can be found on NASA's MAST website.

\section*{Acknowledgment}

The authors are thankful to the anonymous reviewer for providing insightful comments that significantly improved the manuscript substantially. For this work, MS acknowledges Dr. Simon J. Murphy for helping with the ideas. The work presented here is supported by the Belgo-Indian Network for Astronomy and Astrophysics (BINA), sanctioned by the Department of Science and Technology (DST, Government of India; DST/INT/Belg/P-09/2017) and the Belgian Federal Science Policy Office (BELSPO, Government of Belgium; BL/33/IN12). MS thanks CSIR for supporting the fellowship (09/948(0006)/2020-EMR-1). SJ and OT acknowledge the financial support received from the BRICS grant DST/ICD/BRICS/Call-5/SAPTARISI/2023(G). OT acknowledges the International Science Programme (ISP) of Uppsala University for financial support. IY is grateful to the Russian Science Foundation (RSCF grant 21-12-00147) for the partial financial support. 
AG, DM, and SJ are grateful for the support received from the Indo-Thailand Programme of Co-operation in Science and Technology through the Indo-Thai joint project DST/INT/Thai/P-16/2019.    The Python modules were used in our analysis are \texttt{lightkurve, astropy, matplotlib, scipy, numpy}.




\bibliographystyle{mnras}
\bibliography{The_pulsating_variable_HD_118660} 



\appendix


\bsp	
\label{lastpage}
\end{document}